\documentstyle[12pt,epsfig]{article}
\newcommand{\beq}{\begin{equation}}
\newcommand{\eeq}{\end{equation}}
\newcommand{\bea}{\begin{eqnarray}}
\newcommand{\eea}{\end{eqnarray}}
\newcommand{\eps}{\epsilon}
\newcommand{\nn}{\nonumber}
\newcommand{\epsp}{\epsilon^\prime}
\newcommand{\epsi}{\frac{\epsilon^\prime}{\epsilon}}
\newcommand{\epsil}{\epsilon^\prime/\epsilon}
\newcommand{\DSone}{\Delta S=1}
\newcommand{\DBtwo}{\Delta B=2}
\newcommand{\tsp}{\rule{0pt}{2.1ex}}
\newcommand{\tsq}{\rule{0pt}{2.6ex}}

\begin{document}
\topmargin -0.6cm
\oddsidemargin -0.5cm
\evensidemargin -0.5cm
\pagestyle{empty} 
\begin{flushright}
CERN-TH/97-2\\
\end{flushright}
\vspace*{20mm}
\begin{center}
{\Large\bf \boldmath$\frac{\epsilon^\prime}{\epsilon}$: three years
later}\footnote{
Invited talk to the 4th KEK Topical Conference on ``Flavor Physics'', KEK,
Japan, October 1996}
\end{center}
\vspace{1.5cm}

\begin{center}
{\bf M.~Ciuchini}\footnote{On leave of absence from INFN,
Sezione Sanit\`a, V.le Regina Elena 299, 00161 Rome, Italy}\\
{\small Theory Division, CERN, 1211 Geneva 23, Switzerland}
\end{center}
\vspace{3.5cm}
\centerline{\bf Abstract}

Three years after the completion of the next-to-leading order calculation,
the status of the theoretical estimates of $\epsil$ is reviewed. In spite of
the theoretical progress, the prediction of $\epsil$ is still affected by a
$100\%$ theoretical error.
In this paper the different sources of uncertainty are
critically analysed and an updated estimate of $\epsil$ is presented.
Some theoretical implications of a value of $\epsil$ definitely larger than
$10^{-3}$ are also discussed. 

\vfill
\begin{flushleft}
CERN-TH/97-2\\
January 1997
\end{flushleft}
\eject
~\vfill\eject

% put your own definitions here:
%   \newcommand{\cZ}{\cal{Z}}
%   \newtheorem{def}{Definition}[section]
%   ...

% declarations for front matter

\setcounter{page}{1}
\setcounter{footnote}{0}
\baselineskip20pt
\pagestyle{plain}

\section{Introduction}

After many years, the direct $CP$ violation in $K^0$ decays, 
parametrized by $\epsp$, is still an open issue.
The last generation experiments have found \cite{e731,na31}
\beq
\epsi=\left\{
\begin{tabular}{l l}
$(7.4\pm 5.9)\times 10^{-4}$ & E731\\
$(20\pm 7)\times 10^{-4}$ & NA31
\end{tabular}\right.
\eeq
From these results no definite conclusions can be drawn on the
$CP$ property of the $K^0$ decay vertices, namely on whether $\epsp$ is
vanishing.

Theoretically, the Standard Model makes precise assumptions on the
mechanism that generates the $CP$ violation. Indeed the only source of
$CP$ violation is the free phase which appears in the
Cabibbo-Kobayashi-Maskawa
(CKM) matrix \cite{ckm} with three quark generations. This choice implies a
non-vanishing $\epsp$, unless some dynamical cancellation occurs. The
$CP$-violating phase appears in the $K^0$ decays through the
so-called penguin diagrams. However other choices are possible, for
example the superweak model \cite{wo64}, which predicts strictly
$\epsp=0$.

To clarify this issue, a new generation of experiments is going to
be built, achieving a sensitivity on $\epsil$ at the level of
$1$--$2\times 10^{-4}$ \cite{newexp}. On the theoretical side, the
problem is giving a reliable estimate of $\epsil$, including the
theoretical error. The task is not easy: physics from many scales
effectively contribute to $\epsil$, from the top mass down to the strange
mass, including important non-perturbative effects. Nevertheless all
recent analyses agree on predicting $\epsil=\mbox{few}\times 10^{-4}$,
with roughly a $100\%$ relative error \cite{epsith,cfmrs95}.

In the following we present an updated prediction of $\epsil$, giving an
account of the procedure and the different sources of theoretical
uncertainty. We also discuss the dependence of $\epsil$ on some critical
non-perturbative parameters.

\section{\boldmath$\epsil$ in a few steps}

The essential theoretical tool for the calculation of $\epsil$ is the 
$\DSone$ effective Hamiltonian, which allows the separation of the
short- and long-distance physics.
Using the effective Hamiltonian, one obtains
an expression of $\epsil$ that involves CKM parameters, Wilson coefficients
and local operator matrix elements.
Therefore the evaluation of $\epsil$ requires essentially three steps,
namely $(1)$ the phenomenological determination of the CKM parameters,
$(2)$ the calculation of the Wilson coefficients at a next-to-leading order
(NLO) and $(3)$ the determination of the matrix elements of the local
operators appearing in the $\DSone$ effective Hamiltonian. 

\subsection{Basic formulae}

The NLO $\DSone$ effective Hamiltonian at a scale $m_b > \mu > m_c$ can
be written as
\bea
{\cal H} &=&-\frac {\lambda_u G_F} {\sqrt{2}}
\Bigl\{ (1 - \tau ) \Bigl[ C_1(\mu)\Bigl( Q_1(\mu)- Q_1^c(\mu) \Bigr)
+ C_2(\mu)\left( Q_2(\mu) - Q_2^c(\mu) \right)  \Bigr]\nn\\
&~&+ \tau \sum_{i=1}^{9} C_i(\mu) Q_i(\mu) \Bigr\}\,~,
\label{eq:heff}
\eea
where $G_F$ is the Fermi constant, $\lambda_q=V_{qd} V^\star_{qs}$ and
$\tau=-\lambda_t/\lambda_u$, $V_{q_i q_j}$ being the CKM matrix elements.
The $CP$-conserving and $CP$-violating contributions are easily separated,
the latter being proportional to $\tau$.

The operator basis includes eleven dimension-six local four-fermion
operators\footnote{One more operator must be included if $\mu > m_b$
is considered.}. They are given by
\bea
Q_{ 1} &=&({\bar s}_{\alpha}d_{\alpha})_{ (V-A)}
    ({\bar u}_{\beta}u_{\beta})_{(V-A)}
   \nn\\
Q_{ 2} &=&({\bar s}_{\alpha}d_{\beta})_{ (V-A)}
    ({\bar u}_{\beta}u_{\alpha})_{ (V-A)}
\nn \\
Q_{ 1}^c &=&({\bar s}_{\alpha}d_{\alpha})_{ (V-A)}
    ({\bar c}_{\beta}c_{\beta})_{(V-A)}
   \nn\\
Q_{ 2}^c &=&({\bar s}_{\alpha}d_{\beta})_{ (V-A)}
    ({\bar c}_{\beta}c_{\alpha})_{ (V-A)}
\nn \\
Q_{3,5} &=&
    ({\bar s}_{\alpha}d_{\alpha})_{ (V-A)}
    \sum_q({\bar q}_{\beta}q_{\beta})_{ (V\mp A)}
\nn\\
Q_{4,6} &=& ({\bar s}_{\alpha}d_{\beta})_{ (V-A)}
    \sum_q({\bar q}_{\beta}q_{\alpha})_{ (V\mp A)}
 \\
Q_{7,9} &=& \frac{3}{2}({\bar s}_{\alpha}d_{\alpha})_
    { (V-A)}\sum_{q}e_{ q}({\bar q}_{\beta}q_{\beta})_
    { (V\pm A)}
\nn \\
Q_{8} &=& \frac{3}{2}({\bar s}_{\alpha}d_{\beta})_
    { (V-A)}\sum_{q}e_{ q}({\bar q}_{\beta}q_{\alpha})_
    { (V+ A)}~,\nn
\label{eq:opbasis}
\eea
where $(\bar q_\alpha q^\prime_\beta)_{(V\pm A)}=\bar q_\alpha\gamma_\mu
(1\pm\gamma_5)q^\prime_\beta$, $\alpha$ and $\beta$ are colour indices,
and the sum index $q$ runs over $\{d,u,s,c\}$.
Operators $Q_3$--$Q_6$ are generated by the insertion of the tree level
operator $Q_2$ into the strong penguin diagram, while $Q_7$--$Q_9$ come
from the electromagnetic penguin diagrams. As we will see, the two
classes of operators are both relevant for $\epsil$.
Further details on the NLO $\DSone$ effective Hamiltonian can be found
in ref. \cite{ds1}

From the definition of $\epsp$ and using eq. (\ref{eq:heff}), one
readily obtains
\beq
\eps^\prime=i\frac{e^{ i(\delta_2-\delta_0)}}{\sqrt{2}}\frac{\omega}
{\mbox{Re}A_{ 0}}\Bigl[\omega^{ -1}
(\mbox{Im}A_{ 2})^{\prime}
-(1-\Omega_{ IB})\,\mbox{Im}A_{ 0}\Bigr]~,
\label{eq:epsp}
\eeq
where, as usual, $A_I e^{i\delta_I}=\langle \pi\pi(I)\vert {\cal H} \vert
K^0\rangle$, $\Omega_{IB}$ is the isospin breaking contribution due to the
$\pi$--$\eta$--$\eta^\prime$ mixing \cite{bg87}, $\omega=\mbox{Re}A_2/
\mbox{Re}A_0$, and
\bea
\mbox{Im}A_{ 0} &=&-G_F \mbox{Im}\Bigl({ V}_{ ts}^{ *}{ V}_{ td}\Bigr)
\Bigl\{-\Bigl(C_{ 6}B_{ 6}
+\frac{1}{3}C_{ 5}B_{ 5}\Bigr)Z
+\Bigl(C_{ 4}B_{ 4}+\frac{1}{3}C_{ 3}B_{ 3}\Bigr)X\nn\\
&&+C_{ 7}B_{ 7}^{ 1/2}\Bigl(\frac{2Y}{3}+\frac{Z}{6}+
\frac{X}{2}\Bigr)
+C_{ 8}B_{ 8}^{ 1/2}\Bigl(2Y+\frac{Z}{2}+\frac{X}{6}\Bigr)\nn\\
&&-C_{ 9}B_{ 9}^{ 1/2}\frac{X}{3}
+\Bigl(\frac{C_{ 1} B_{ 1}^{ c}}{3}+C_{ 2}B_{ 2}^{ c}\Bigr)X\Bigr\}~,\nn\\
\\
(\mbox{Im}A_{ 2})^{\prime}\!&=&\!-G_F\mbox{Im}\Bigl({ V}_{ ts}^{ *}{ V}_{
td}\Bigr)
\Bigl\{C_{ 7}B_{ 7}^{ 3/2}\Bigl(\frac{Y}{3}
-\frac{X}{2}\Bigr)+C_{ 8}B_{ 8}^{ 3/2}\Bigl(Y-\frac{X}{6}\Bigr)\nn\\
& & +C_{ 9}B_{ 9}^{ 3/2}\frac{2X}{3}\Bigr\}~.\nn
\label{eq:ima0a2}
\eea
The relevant operator matrix elements are given in terms of the
$B$-parameters as follows:
\bea
\langle \pi\pi(0)\vert Q_i\vert K\rangle &=&
B^{1/2}_i\langle \pi\pi(0)\vert Q_i\vert K\rangle _{VIA}\nn\\
\langle \pi\pi(2)\vert Q_i\vert K\rangle &=&
B^{3/2}_i\langle \pi\pi(2)\vert Q_i\vert K\rangle _{VIA}~,
\eea
where the subscript $VIA$ means that the matrix elements are calculated in
the vacuum insertion approximation. VIA matrix elements can be calculated
and expressed in terms of the three quantities
\bea
X\!&=&\!f_{\pi}\left(M_{ K}^{ 2}-M_{\pi}^{ 2}\right)~,\nn\\
Y\!&=&\!f_{\pi}\left(\frac{M_{ K}^{ 2}}{m_s(\mu)+m_d(\mu)}\right)^2
 \sim 12\,X\left(\frac{150 \, \mbox{MeV}}{m_s(\mu)}\right)^2~,\\
Z\!&=&\!4\left(\frac{f_{ K}}{f_{\pi}}-1\right)Y~.\nn
\label{eq:xyz}
\eea
Notice that, contrary to $X$ and $Z$, $Y$ does not vanish in the chiral
limit. This reflects the different chiral properties of the operators
$Q_7$ and $Q_8$.

\subsection{CKM matrix elements}

\begin{figure*}[t]
\centering
\epsfysize=0.5\textheight
\epsfxsize=\textwidth
\epsffile{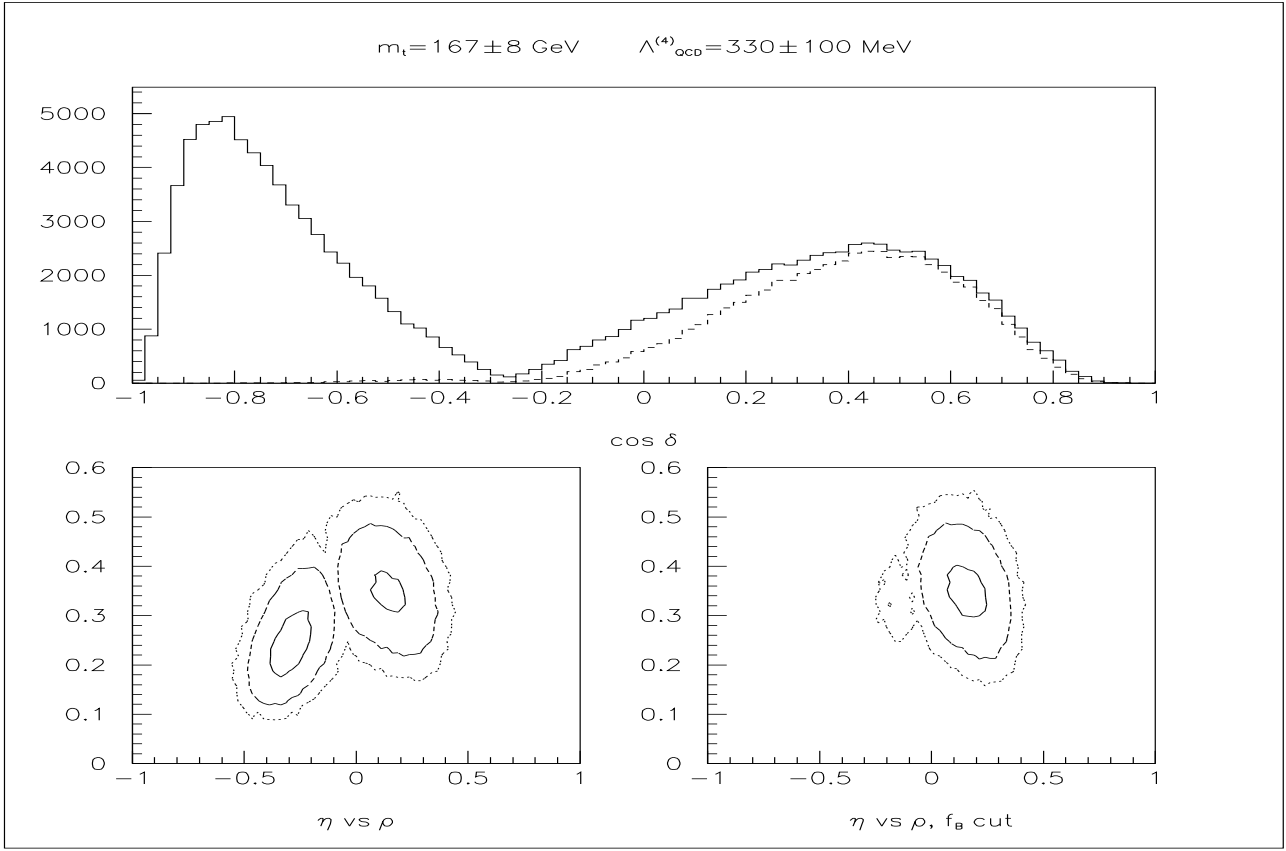}
\caption[]{\it{Constraints on the $C$P-violating phase and the
unitarity triangle. The $\cos\delta$ distributions are plotted above.
The dashed distribution includes the constraint coming from
$\Delta M_{B_d}$ and the lattice determination of $B_B\sqrt{f_b}$.
Below, the same constraints are shown as contour plots in the
$\rho$--$\eta$ plain. The solid, dashed and dotted lines correspond
to the $5\%$, $68\%$ and $95\%$ of the generated configurations
respectively. In this plane the allowed region determines the
third vertex of the unitarity triangle $\sum_{q=\{u,c,t\}}
V^\star_{qb} V_{qd}=0$, the others being $(0,0)$ and $(1,0)$.
}}
\label{fig:ckm}
\end{figure*}

\begin{table}
\centering
\begin{tabular}{|c|c|}\hline
\tsp $G_F$ & $1.16639\times 10^{-5}\,\mbox{GeV}^{-2}$ \\[0.1ex]\hline
\tsp $m_c$ & $1.5$ GeV \\[0.1ex]
\tsp $m_b$ & $4.5$ GeV \\[0.1ex]\hline
\tsp $M_W$ & $80.32\,\mbox{GeV}$ \\[0.1ex]
\tsp $M_B$ & $5.279\,\mbox{GeV}$\\[0.1ex]
\tsp $M_K$ & $498\,\mbox{MeV}$ \\[0.1ex]
\tsp $\Delta M_K$ & $3.495\times 10^{-12}\,\mbox{MeV}$ \\[0.1ex]
\tsp $M_\pi$ & $140\,\mbox{MeV}$\\[0.1ex]\hline
\tsp $f_\pi$ & $132\,\mbox{MeV}$\\[0.1ex]
\tsp $f_K$ & $160\,\mbox{MeV}$\\[0.1ex]\hline
\tsp $\mbox{Re}A_0$ & $2.7\times10^{-7}\mbox{GeV}$ \\[0.1ex]
\tsp $\omega$ & $0.045$\\[0.1ex]\hline
\tsp $\lambda=\sin\theta_c$ & $0.221$\\[0.1ex]\hline
\tsp $\epsilon^{exp}$ & $2.284\times 10^{-3}$ \\[0.1ex]\hline
\tsp $\mu$ & $2\,\mbox{GeV}$\\[0.1ex]
\hline
\end{tabular}
\caption[]{\it Input parameters assumed to be constants in the analysis}
\label{tab:const}
\end{table}

In order to estimate $\epsil$, we need $\mbox{Im}\Bigl({ V}_{ ts}^\star{
V}_{td}\Bigr)$. This requires a complete knowledge of the CKM matrix $V$.
For example, in the Wolfenstein parametrization \cite{wo83}, up to
$O(\lambda^3)$,
\beq
V=
\pmatrix{1-\frac{\lambda^2}{2}&\lambda&A\lambda^3\left(\rho-i\eta\right)\cr
-\lambda & 1-\frac{\lambda^2}{2} & A\lambda^2\cr
A\lambda^3\left(1-\rho+i\eta\right)&-A\lambda^2&1\cr}\nn
\eeq
and
\beq
\mbox{Im}\Bigl({ V}_{ ts}^\star{ V}_{td}\Bigr)=-A^2\lambda^5\eta=
-A^2\lambda^5\sigma\sin\delta~,
\eeq
using the definition of the $CP$-violating phase $\delta$ given by
$\sigma e^{i\delta}=\rho+i\eta$. While $\lambda$ is well known, $A$ and
$\sigma$
can be extracted from the measurements of the $B$ lifetime and the semileptonic
decay rates, respectively. The remaining task is the determination of
the $CP$-violating phase. The $CP$-violating parameter $\eps$ in the
$K^0$--$\bar K^0$ mixing is the obvious tool. It is given by
\beq
\vert\epsilon\vert_{\xi=0}=C_{ \epsilon}B_{ K}A^2\lambda^6\sigma\sin\delta
\left\{F(x_c,x_t)+
F(x_t)[A^2\lambda^4(1-\sigma\cos\delta)]-F(x_c)\right\}\, ,\nn
\label{eq:eps}
\eeq
where $F(x_i)$ and $F(x_i,x_j)$ are the Inami-Lim functions \cite{il81},
including the QCD corrections \cite{ds2}, and
\beq
C_{ \epsilon}=\frac
{G_{ F}^2f_{ K}^2 M_{ K}M_{ W}^2}{6\sqrt 2{\pi}^2\Delta M_K}\, .
\eeq
The comparison of the previous expressions with the measured values of $\eps$
allows the extraction of $\cos\delta$. However, since eq. (\ref{eq:eps})
is a quadratic function of $\cos\delta$, one obtains two different
solutions, corresponding roughly to $\cos\delta$ positive and negative.
The solid-line
distribution of $\cos\delta$ in fig.~\ref{fig:ckm} is the result of this
analysis. The distribution is obtained by generating randomly the
various parameters according to the values and errors listed in
tables \ref{tab:const} to \ref{tab:varg}.

\begin{table}
\centering
\begin{tabular}{|c|c|}\hline
\tsp $\Lambda^{(4)}_{QCD}$ & $330\pm 100\,\mbox{MeV}$\\[0.1ex]\hline
\tsp $\Omega_{IB}$ & $0.25\pm 0.10$\\[0.1ex]\hline
\tsp $(f_B B_B^{1/2})_{th}$ & $210\pm 35$ MeV\\[0.1ex] \hline
\tsq $B_{K}^{RG-inv}$ & $0.75\pm 0.15$\\[0.1ex]
\tsp $B_{1-2}^c$ & $0 - 0.15^{(*)}$\\[0.1ex]
\tsp $B_{3,4}$ & $1 - 6^{(*)}$\\[0.1ex]
\tsp $B_{5,6}$ & $1.0\pm 0.2$\\[0.1ex]
\tsp $B_{7-8-9}^{(1/2)}$ & $1^{(*)}$\\[0.1ex]
\tsp $B_{7}^{(3/2)}$ & $0.6\pm 0.1$\\[0.1ex]
\tsp $B_{8}^{(3/2)}$ & $0.8\pm 0.15$\\[0.1ex]
\tsp $B_9^{(3/2)}$ & $0.62\pm 0.10$\\[0.1ex]
\hline
\end{tabular}
\caption[]{\it{Input parameters assumed to be uniform in the analysis}}
\label{tab:varl}
\end{table}

To further constrain the $CP$-violating phase, one can exploit the
$B_d-\bar B_d$ mass difference, given by
\beq
\Delta M_{B_d}=\frac{G_F^2 M_W^2 M_B^2}{6 \pi^2}\frac{f_B^2}{M_B} B_B
A^2 \lambda^6 \Bigl( 1 +\sigma^2
-2 \sigma \cos\delta \Bigr) F(x_t)~,
\label{eq:xd}
\eeq
where $B_B$ is the $B$-parameter associated to the $\DBtwo$ operator
$(\bar b d)_{(V-A)} (\bar b d)_{(V-A)}$. Recent lattice results
give quite large values of $B_B \sqrt{f_B}$ \cite{fblat}. We use
$B_B \sqrt{f_B}=210\pm 35~\mbox{MeV}$.
The requirement of compatibility between $B_B\sqrt{f_B}$ extracted from
eq.~(\ref{eq:xd}) and the lattice value results in an effective 
selection of the positive values of $\cos\delta$; see the dashed
$\cos\delta$ distribution in fig.~\ref{fig:ckm}. Therefore we can quote the
value
\beq
\cos\delta=0.38\pm 0.23~,
\eeq
where the error is the variance of the dashed distribution in
fig.~\ref{fig:ckm}. 

\subsection{Wilson coefficients}

The Wilson coefficients $C_i(\mu)$ appearing in eq.~(\ref{eq:ima0a2}) can
be calculated using the renormalization group improved perturbation theory,
provided that $\mu$ is a scale large enough for the perturbation theory
to be reliable. Indeed their $\mu$ dependence is controlled by the
renormalization group equation
\beq
\mu^2\frac{d}{d\mu^2}C_i(\mu)=\frac{1}{2}\sum_j\hat\gamma_{ji}C_j(\mu)~,
\label{eq:rge}
\eeq
where $\hat\gamma$ is the anomalous dimension matrix of the operators in
eq.~(\ref{eq:opbasis}). The NLO calculation of these
coefficients is discussed in ref.~\cite{ds1}. To our end, it
suffices to recall that, for any suitable value of $\mu$, the Wilson
coefficients are a known set of real numbers, which however still depend
on the scheme chosen to renormalize the local operators.
The typical relative error on the coefficients relevant for $\epsil$ is
$10$--$20\%$. For more details
and numerical values, see for example ref.~\cite{cfmrs95}.

\begin{table}
\centering
\begin{tabular}{|c|c|}\hline
\tsp $m_t^{\overline{MS}}(m_t)$ & $167\pm 8\,\mbox{GeV}$\\[0.1ex]\hline
\tsp $|V_{cb}|=A\lambda^2$ & $0.040\pm 0.003$\\[0.1ex]
\tsp $|V_{ub}/V_{cb}|=\lambda\sigma$ & $0.080\pm 0.015$\\[0.1ex] \hline
\tsp $\tau_B$ & $1.56\pm 0.06\, \mbox{ps}$\\[0.1ex]\hline
\tsp $\Delta M_{B_d}^{exp}$ & $0.464\pm 0.018\, \mbox{ps}^{-1}$\\[0.1ex]\hline
\tsq $m_s^{\overline{MS}}(2\,\mbox{GeV})$ & $128\pm 18$ MeV\\[0.1ex]
\hline
\end{tabular}
\caption[]{\it{Input parameters assumed to be Gaussian in the analysis}}
\label{tab:varg}
\end{table}

\subsection{Local operator matrix elements}

The calculation of the matrix elements of the local operators appearing
in eq.~(\ref{eq:heff}) requires the use of a non-perturbative technique.
Indeed these matrix elements contain the low-energy QCD dynamics, from the
scale $\mu$ downward. Besides the $\mu$ dependence, at the NLO
they also depend on the operator regularization scheme. Both these
dependences must be matched with the corresponding dependence in the Wilson
coefficients, in order to have a scale- and scheme-independent physical
prediction. Therefore only a non-perturbative approach which allows a full
control over the renormalization scale and scheme dependences can be
consistently used. Furthermore this technique should allow choosing the
renormalization scale $\mu$ large enough for the perturbative calculation
of the Wilson coefficients to be reliable.
As far as $\epsil$ is concerned, the only known non-perturbative approach
that fulfills these requirements is lattice QCD. Several $B$-parameters
appearing in eq.~(\ref{eq:ima0a2}) have been computed on the lattice
\cite{lattice}. In particular $B_6$ and $B_8^{(3/2)}$, which turn out to be
the most important numerically, are known. For the others, we use the
vacuum insertion approximation, namely $B=1$ at $\mu=2~\mbox{GeV}$.
There are two exceptions:
first, $B_3$ and $B_4$ are allowed to be as large as 6, considering that the
penguin operator matrix elements may be at least partially responsible for the
$\Delta I=1/2$ rule. The other is $B_{1,2}^c$, which strictly speaking
cannot even be defined, since $\langle\pi\pi\vert Q^c_{1,2}\vert K^0
\rangle_{VIA}=0$. However, a small contribution is expected beyond the
VIA and we parametrize it by assuming the
VIA matrix elements to be equal to those of $Q_{1,2}$ and introducing a
small $B$-parameter. Table~\ref{tab:varl} contains the values of the
$B$-parameters used in the numerical analysis.

The relevant $B$-parameters are known with an error of about $20\%$.
We will see that unfortunately they produce a
larger error in $\epsil$ because of a partial cancellation between different
terms. However,
there is another source of uncertainty stemming from the normalization
of the matrix elements, namely the value of the running strange mass,
see eqs.~(\ref{eq:ima0a2})--(\ref{eq:xyz}). The former good agreement between
lattice and QCD sum rules on this mass \cite{all94,jm95} has been
questioned by a recent lattice result, in which the extrapolation to the
continuum limit is attempted \cite{gb96}. We will quantitatively discuss this
issue in the
next section; however, it is worth noting that this uncertainty comes only
from the choice of normalizing the lattice results to the vacuum insertion
approximation. In the future the possibility of normalizing lattice results
to a better known quantity should be considered.

\subsection{The ``best'' estimate}

\begin{figure*}[t]
\centering
\epsfysize=0.5\textheight
\epsfxsize=\textwidth
\epsffile{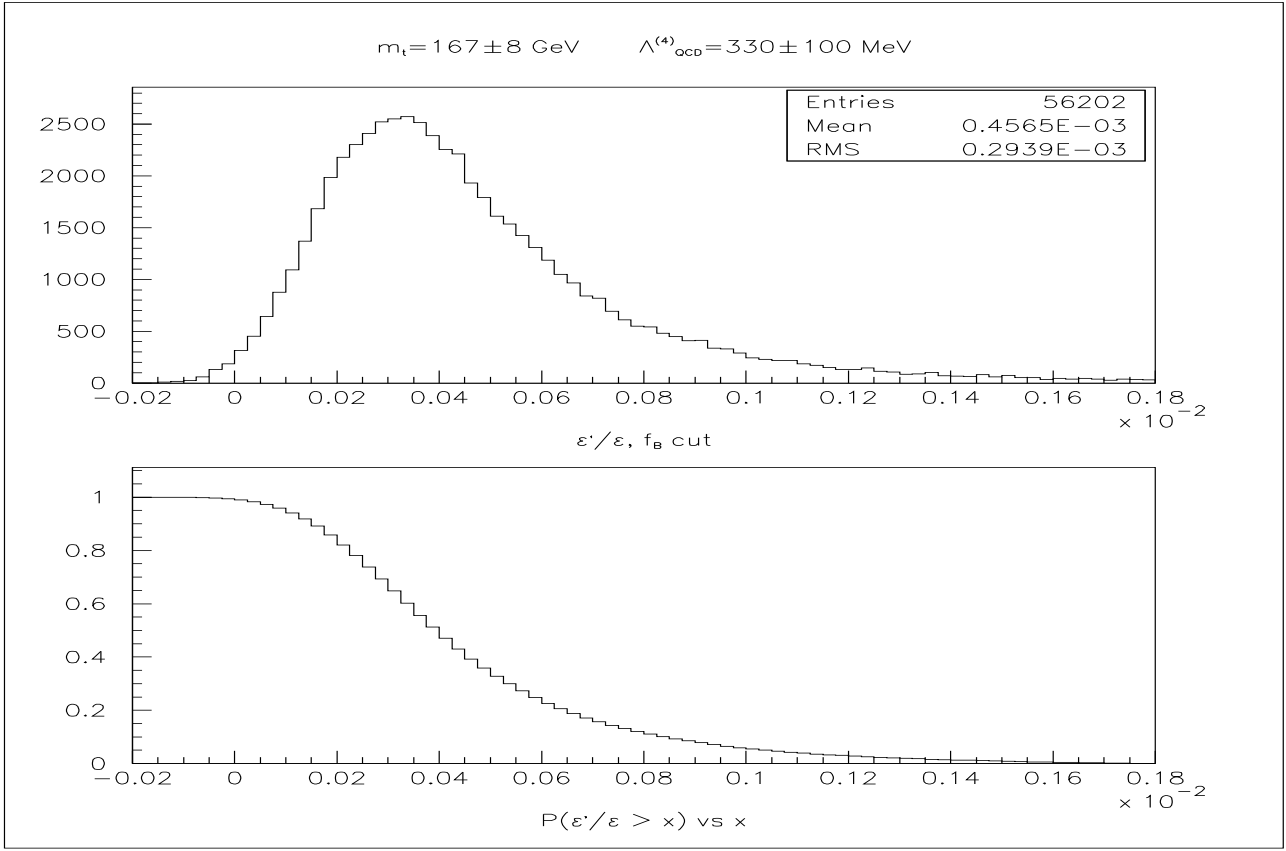}
\caption[]{\it{Plot of the distribution of $\epsil$, $p(\epsil)$, above, 
and of $P(\epsil>x)=\int_x^\infty dy\,p(y)/\int_{-\infty}^\infty dy\,p(y)$,
below
}}
\label{fig:epsi}
\end{figure*}

Having the necessary ingredients, we can put them togheter and produce an
estimate of $\epsil$. Varying randomly the parameters in
tables \ref{tab:varl} and \ref{tab:varg}, we obtain the distribution of
fig.~\ref{fig:epsi}, from which we estimate
\beq
\epsi=(4.6 \pm 3.0\pm 0.4)\times 10^{-4}~.
\label{eq:epsires}
\eeq
Again the first error is the variance of the distribution, while the second
one refers to the residual scheme dependence coming from
higher order in perturbation theory and it is obtained by using two different
renormalization schemes.

Figure~\ref{fig:epsi} also shows the probability distribution $P(\epsil
> x)$ as a function of $x$. For instance, given our choice of the input
parameters and distributions, one obtains $P(\epsil
> 2\times 10^{-4})\sim 0.8$.

\begin{figure}[t]
\centering
\epsfysize=0.4\textheight
\epsfxsize=\textwidth
\epsffile{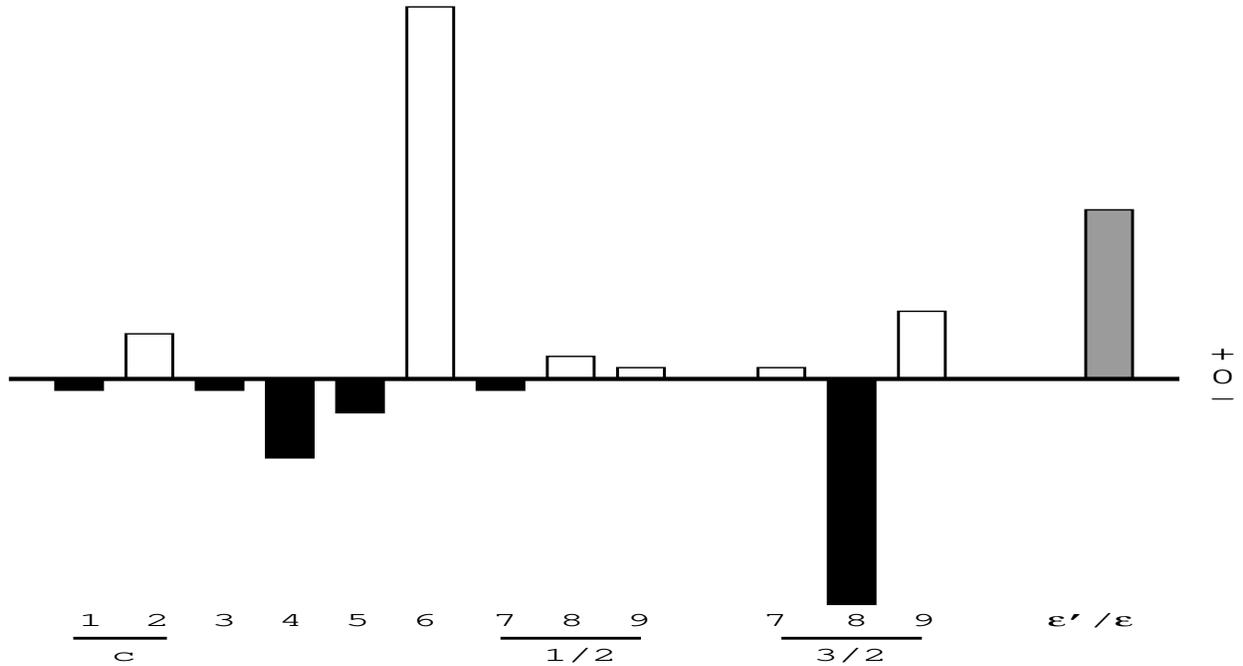}
\caption[]{\it{Relative contributions of the different terms appearing
in eq.~(\ref{eq:ima0a2}) with respect to $\epsil$. The largest terms
are those containing $B_6$ and $B_8^{(3/2)}$. A partial cancellation
occurs between them.}}
\label{fig:contribs}
\end{figure}

Why is the theoretical error $100\%$ or even larger? Wilson
coefficients and $B$-parameters have errors $\sim 20\%$ or less. One can
argue that there are many contributions, but fig.~\ref{fig:contribs} shows
that actually the bulk of the result can be obtained retaining only two
operators. On the other hand the two main terms, say $B_6$ and
$B_8^{3/2}$, partially cancel each other, lowering the central value of the
prediction and increasing the relative error. The effectiveness of this
cancellation depends not only on the top mass, which nowadays is a well-known
quantity, but also on the isospin-breaking parameter $\Omega_{IB}$, of which
we have only quite old theoretical estimates \cite{bg87}. Contributions to the
error also come from some overall factors, $\sin\delta$ and
$m_s^{-2}$ which appears in front of the largest terms. All these effects
sum up to give the large error of the final result.

\section{The future of \boldmath$\epsil$}
Given the theoretical difficulties in reducing the error on $\epsil$, we
may wonder what we will learn from the next generation of experiments, with
an expected sensitivity at the level of $2\times 10^{-4}$.

\subsection{What the present estimate is good for}

According to present theoretical estimates, the Standard Model can accomodate 
$\epsil$ values between $0$ and $10^{-3}$.
If the present ambiguity on
the experimental value of $\epsil$ will be solved, confirming the
E731 figure, then the new measurement will hardly 
contribute to improving our knowledge of the Standard Model
and the QCD dynamics. Indeed, even if a non-vanishing value of $\epsil$
would rule out the superweak model allowed at present, a measurement of
$\epsil < 10^{-3}$ can be accommodated by the Standard Model with
no pain, weakly constraining the different parameters that enter
the theoretical expression of $\epsil$.

On the contrary a measurement of $\epsil > 10^{-3}$ would be a fruitful 
surprise. Its explanation would require either new physics at work or
an improved understanding of the QCD dynamics. 

\subsection{What does \boldmath$\epsil > 10^{-3}$ imply?}

In the following let us assume that indeed the future measurements of
$\epsil$ give a value definitely larger than $10^{-3}$. New sources
of $CP$ violation beyond the Standard Model can change the prediction
in eq.~(\ref{eq:epsires}) and accommodate such a large value. However
the task is not trivial. For instance supersymmetric effects on $\epsil$ 
are small in comparison with the large error, in the Minimal Supersymmetric
Standard Model \cite{gg95}. 

Anyway, let us ignore here the possibility of new physics effects and
stick to the Standard Model. One or more input parameters must deviate
from the values listed in tables \ref{tab:const}--\ref{tab:varg} to
account for the increase of $\epsil$ in this case.
We believe that these tables collect the ``best''
set of input parameters allowed by our present knowledge of the
CKM matrix and the QCD dynamics. However, one can still consider the
effect on $\epsil$ caused by assuming that some critical parameter is
outside its allowed range. This is a particularly reasonable speculation
in the case of the $B$-parameters and the hadronic quantities in general.
Indeed the systematic errors of the lattice QCD results are not completely
under control at present and may be underestimated in some cases.
Hopefully new developments of the lattice QCD techniques, such as 
non-perturbative renormalization and unquenching, will
lead to a better control of the systematic errors in the future.

Let us consider two scenarios leading to a large $\epsil$: first a small
strange mass, then anomalous $B_6$ and $B_8^{3/2}$.

The possibility of a small strange mass, at the level of
$50$--$90~\mbox{MeV}$, has been suggested by a recent compilation of
lattice results \cite{gb96}. A previous lattice analysis, done at finite
lattice spacing, gave a running mass
$m_s^{\overline{MS}}(2~\mbox{GeV})=128\pm 18~\mbox{MeV}$ \cite{all94}, in
agreement with the QCD sum rules determination \cite{jm95}.
The new analysis of ref.~\cite{gb96} extrapolates the value of the strange
mass to zero lattice spacing, finding $m_s^{\overline{MS}}
(2~\mbox{GeV})=90\pm 15~\mbox{MeV}$ in the quenched approximation. The
inclusion of dynamical fermions further decreases $m_s$
and they quote $m_s^{\overline{MS}}(2~\mbox{GeV})=70\pm 11~\mbox{MeV}$
for two generations of dynamical fermions.
This result indicates at least that the error associated to the finite
lattice spacing was probably underestimated in the previous analysis.
However, we believe that it is premature to accept the new figures for
the running strange mass, until the discrepancy with the QCD sum rules
result is understood.

Anyway it is an easy exercise to see how $\epsil$ changes in this
small-$m_s$ scenario, since the largest terms are proportional to $m_s^{-2}$.
We obtain
\beq
\epsi=(14\pm 8)\times 10^{-4},
\eeq
for $m_s^{\overline{MS}}(2~\mbox{GeV})=70\pm 11~\mbox{MeV}$.
Thus a small strange mass can push $\epsil$ up to some units in $10^{-3}$.
Notice however that the relative error is roughly unchanged and a vanishing
$\epsil$ is still well inside the allowed range. We stress again that, as
far as $\epsil$ is concerned, the problem of $m_s$ appears as the
consequence of an inappropriate choice of the $B$-parameter normalization.

In another conceivable scenario, one can take $B_6$ to be large and  
at the same time a small $B_8^{3/2}$. This choice tends to spoil the
cancellation between strong- and electro-penguin operators, thus increasing
$\epsil$. Indeed a recent
result gives a new value $B_8^{3/2}=0.81\pm 0.03^{+0.03}_{-0.02}$ smaller 
than the previous
ones \cite{gb96,gbs96}. To our knowledge, no new results are available for
$B_6$. We can explore an extreme scenario where we choose
\beq
B_6=1.50\pm 0.15~\mbox{and}~B_8^{3/2}=0.50\pm 0.05~.
\eeq
In this case, we obtain
\beq
\epsi=(12\pm 4)\times 10^{-4}
\eeq
Again $\epsil$ larger than $10^{-3}$ is predicted. The relative error is
smaller than before because of the spoiling of the cancellation between
$B_6$ and $B_8^{(3/2)}$.
Now $\epsp=0$ is only marginally compatible with the prediction.
Notice, however, that precisely the very peculiar situation of a large $B_6$
and a small $B_8^{(3/2)}$ must be realized in order to enhance $\epsil$.

The previous examples show that a measurement of $\epsil$ larger than
$10^{-3}$ would indeed have non-trivial implications, forcing us either to
reconsider some non-perturbative results or to call for new physics. 

\section*{Acknowledgements}

I benefited from comments by E.~Franco, W.J.~Marciano, G.~Martinelli
and J.L.~Rosner.

\end{document}